\documentclass[prl,amsmath,amssymb,showpacs,twocolumn]{revtex4}
 \usepackage{graphicx}
 \usepackage[latin1]{inputenc}
 \usepackage{xspace}
 \usepackage{latexsym}
 \usepackage{amsmath}
 \usepackage{amssymb}
 \usepackage{amstext}
 \usepackage{amsxtra}
 \usepackage{makeidx}

 \newtheorem{theo}{Theorem}

\def\O{\Omega}

\def\r0{\rho_{0}}

\def\ep{\varepsilon}

\def\a0{\alpha_0}
\def\a{\alpha}

\def\la{\lambda}

\def\be{\begin{equation}}
\def\ee{\end{equation}}
\def\beq{\begin{equation}}
\def\eeq{\end{equation}}

\def\O{\Omega}

\def\({\left(}
\def\){\right)}

\def\pa{\partial}

\newcommand{\zzz}{\bf{Z}} 

\begin{document}

\title{Vortex distribution in the Lowest Landau Level}
\author{Amandine Aftalion, Xavier Blanc}
\affiliation{Laboratoire Jacques-Louis Lions, UMR-CNRS 7598,
Universit{\'e} Paris 6, 175 rue du Chevaleret, 75013 Paris, France. }
\author{Francis Nier}
\affiliation{IRMAR, UMR-CNRS 6625,  Universit\'e Rennes 1, 35042
Rennes Cedex, France.}

\date{\today}


 \begin{abstract}
We study the vortex distribution of the wave functions minimizing
the Gross Pitaevskii energy for a fast rotating condensate in the
Lowest Landau Level (LLL): we prove that the minimizer cannot have
a finite number of zeroes thus the lattice is infinite, but not
uniform.
 This uses the explicit expression of the
projector onto the LLL.
 We also show that any slow varying envelope function can be approximated
  in the  LLL by distorting the
lattice. This is used in particular to approximate the inverted
parabola and understand the role of ``invisible'' vortices: the
distortion of the lattice is very small in the Thomas Fermi region
but quite large outside, where the ``invisible" vortices lie.
 \end{abstract}
 \maketitle


The fast rotating regime for a Bose Einstein condensate in a harmonic trap,
 observed experimentally in \cite{Ketterle1,Boulder04,Boulder04bis},
   displays
analogies with type II superconductors behaviors and Quantum Hall
Physics. However, some different features have emerged and are of
interest, in particular due to the existence of a potential
trapping the atoms.

A quantum fluid described by a macroscopic wave function rotates
through the nucleation of quantized vortices \cite{Donnelly91}.
For a condensate confined in a harmonic potential with cylindrical
symmetry around the rotation axis, a limiting regime occurs when
the rotational frequency $\O$ approaches the transverse trapping
frequency: the centrifugal force nearly balances the trapping
force so that the size of the condensate increases and the number
of vortices diverges. The visible vortices arrange themselves in a
triangular Abrikosov lattice. The system is strongly confined
along the axis of rotation, and it is customary to restrict to a
two dimensional analysis in the $x-y$ plane. We will call
$z=x+iy$.
 The hamiltonian  is
similar to that for a charged particle in a magnetic field: for
rotational angular velocities just below the transverse trap
frequency,  the wave function of the condensate can be described
using only  components in the Lowest Landau Level
(LLL):\beq\label{Psi}\Psi(z)=\Phi_0 \prod_{i=1}^N (z-z_i)
e^{-|z|^2/2}\eeq
 where $\Phi_0$ is a normalization factor and the $z_i$ are the
 location of the vortices.
In rescaled units, the reduced energy in the LLL is
\cite{H,WBP,ABD} \beq\label{elll}{\cal E}_{LLL}(\Psi)=\int \Bigl [
(1-\O)|z|^2|\Psi|^2+\frac G2 |\Psi|^4\Bigr ] d^2r\eeq under
$\int\! d^2r|\Psi|^2=1$, where $\O$ is the rotational velocity,
the transverse trap frequency is scaled to 1, and $G$ models the
interaction term: $G=Ng/(d\sqrt{2\pi})$, where $g$ is the two body
interaction strength and $d$ is the characteristic size of the
harmonic oscillator in the direction of the rotation.

 In the
absence of a confining potential, the problem is reduced to the
one studied by Abrikosov \cite{A} for a type II superconductor and
the minimizer is a wave function with a uniform triangular lattice
\cite{K}; its modulus vanishes once in each cell and is periodic
over the lattice. The presence of the confining potential is at
the origin of a slow varying density profile, which can be
described as the mean of the modulus of the wave function on many
cells.
 Ho \cite{H}
predicted that for a uniform lattice, the smoothed density profile
is a gaussian. Various contributions \cite{WBP,CKR,ABD} then
pointed out that the energy can be lowered if this smoothed
density distribution is an inverted parabola rather then a
gaussian. This type of density profile can be achieved either by
taking wave functions with a uniform lattice  but with components
outside the LLL \cite{WBP} or by remaining in the LLL and
distorting the lattice. The study of the distortion has been the
focus of recent papers \cite{CKR,ABD,Macdo} and raises the issue
of the optimal vortex distribution. In the LLL description, there
are two kinds of vortices : the ``visible vortices", which lie in
the region
 where the wave function is
significant (for instance  inside the Thomas Fermi region in the
case of the inverted parabola), and the ``invisible vortices"
which are in the region where the modulus of the wave function is
small.
 The visible vortices form a regular triangular lattice, while the
 invisible ones seem to have a strong distorted shape, whose distribution is
 essential to recreate the inverted parabola profile inside the
  LLL approximation.
These latter are not within reach of experimental evidence, but
can be computed numerically \cite{ABD,Macdo}. An important
theoretical question is the distribution of these invisible
vortices, their number, or an estimate of how many of them are
necessary to approximate the inverted parabola properly inside the
LLL.

Our main result is to prove that in order to minimize the energy
in the LLL, there is a need for an infinite number of vortices.
The main tool that we use is an explicit expression of the
projector onto the LLL. This projector also allows us to
approximate any slow varying density profile by LLL wave
functions.

{\bf Projection onto the LLL and infinite number of zeroes} We
define a small parameter $\ep=\sqrt{1-\O}$ and make the change of
variables $\psi(z)=\sqrt\ep\Psi(\sqrt \ep z)$, so that the
condensate is of size of order 1 and the lattice spacing is
expected to be of order $\sqrt\ep$. The energy gets rescaled as
${\cal E}_{LLL}(\Psi)=\ep E_{LLL}(\psi)$ where
\beq\label{ELLL}E_{LLL}(\psi)=\int  \Bigl [ |z|^2|\psi|^2+\frac G2
|\psi|^4\Bigr ]d^2r.\eeq Moreover, $\psi$ belongs to the LLL so
that $f(z)=\psi(z)e^{|z|^2/2\ep}$ is a holomorphic function and
thus belongs to the so called Fock-Bargman space
\beq\label{fock}{\cal F}=\left\{ f \hbox{ is holomorphic }, \ \int
|f|^2e^{-|z|^2/\ep}d^2r<\infty\right \}.\eeq Let us point out that
such a function $f$ is not only determined by its zeroes and
normalization factor as in (\ref{Psi}), but also by a globally
 defined phase, which is a holomorphic function.
 The  space ${\cal F}$ is a Hilbert space endowed with the scalar
product $\left <f,g\right
> =\int  f(z)g(z)e^{-|z|^2/\ep}d^2r.$
The point of considering this space is that the projection of a
general function $g(z,\bar z)$ onto ${\cal F}$ is explicit, and
called the Szego projector \cite{M,F} :\beq\label{pi}\Pi(g)=\frac
1{\pi\ep}\int
e^{\frac{z\overline{z'}}{\ep}}e^{-\frac{|z'|^{2}}{\ep}}g(z',\bar{
z'})d^2r'.\eeq If $g$ is a holomorphic function, then an
integration by part yields $\Pi(g)=g$.

If one considers the minimization of $E_{LLL}(\psi)$ without the
holomorphic constraint on $f$, then the minimization process
yields that
$|z|^2+G|\psi|^2-\mu=0$, where $\mu$ is the chemical potential due
to the constraint $\int|\psi|^2=1$, so that
 $|\psi|$ is the inverted parabola \beq\label{ip}
\left|\psi\right|^{2}(z)=\frac{2}{\pi
  R^{2}}\left (1-\frac{|z|^{2}}{R^{2}}\right )\!1_{\left\{|z|\leq
  R\right\}},
  R=\sqrt{\mu}=\left(\frac{2G}{\pi}
\right)^{1/4}\!\!\!\!.\eeq The restriction to the LLL prevents
from achieving this specific inverted parabola
 since $\psi e^{|z|^2/2\ep}$ cannot be a holomorphic
 function.
The advantage of the explicit formulation of the projector
$\Pi$ is that it allows us to derive an equation satisfied by
$\psi$ or rather $f$ when minimizing the energy in the LLL. A
proper distribution of zeroes can approximate an inverted parabola
profile but is going to modify the radius $R$ by a coefficient
$b^{1/4}$ coming from the contribution of the vortex lattice to
the energy.
\begin{theo}\label{el}If $f\in{\cal F}$ minimizes \beq
E(f)=\int  \Bigl [ |z|^2|f|^2e^{-|z|^2/\ep}+\frac G2
|f|^4e^{-2|z|^2/\ep}\Bigr ]d^2r\eeq under $\int
|f|^2e^{-|z|^2/\ep}d^2r=1,$ then $f$ is a solution of the
following equation\beq\label{eqf}\Pi\Bigl (
(|z|^2+G|f|^2e^{-|z|^2/\ep}-\mu)f\Bigr )=0\eeq where $\mu$ is the
chemical potential coming from the mass constraint.
\end{theo} Note that given the relation between $f$ and $\psi$,
$E(f)$ and $E_{LLL}(\psi)$ are identical. Equation (\ref{eqf})
comes from the fact that for any $g$ in ${\cal F}$ with $\left
<f,g\right
> =0$, if $f$
minimizes $E$, then we have \beq\label{fgeq}\int \Bigl [ |z|^2\bar
g f e^{-|z|^2/\ep}+\frac G2 |f|^2\bar g fe^{-2|z|^2/\ep}\Bigr
]d^2r=0\eeq and we use the scalar product in ${\cal F}$ and the
definition of the projector to conclude.

The equation for the minimizer allows us to derive that this
minimizer cannot be a polynomial:
\begin{theo}If  $f\in{\cal F}$ minimizes $E$, then $f$ has an infinite
 number of zeroes.\label{nopol}\end{theo}
We are going to argue by contradiction and assume that $f$ is a
polynomial.

 1. The proof first requires another formulation of
(\ref{eqf}). The projector $\Pi$ has many properties \cite{M,ABN}:
in particular, one can check, using an integration by part in the
expression of $\Pi$, that $\Pi_{\ep}(|z|^{2}f)=z\ep\pa_{z}f+\ep
f$. As for the middle term in the equation, one can compute that
if $f$ is a polynomial,
$\Pi_{\ep}\left(e^{-\frac{|z|^{2}}{\ep}}\left|f\right|^{2}
f\right)
=\Pi_{\ep}\left(e^{-\frac{|z|^{2}}{\ep}}\left|f\right|^{2}\right)\Pi_{\ep}
f\\ =
\Pi_{\ep}(\overline{f}(z))\Pi_{\ep}(e^{-\frac{|z|^{2}}{\ep}}f^{2})
=\bar{f}(\ep\pa_{z})\Pi_{\ep}(e^{-\frac{|z|^{2}}{\ep}}f^{2})$. A
simple change of variable yields
$\Pi_{\ep}\left(e^{-\frac{|z|^{2}}{\ep}}f^{2}\right)(z)=(\pi
\ep)^{-2}\int
e^{-\frac{z\overline{z'}-2|z'|^{2}}{\ep}}f^{2}(z')d^2r'\\=\frac 12
\Pi_{\ep}\left(f^2 (\frac {.}{\sqrt 2})\right )(\frac z
{\sqrt{2}})=\frac 12 f^2(\frac z{2})$.  Thus, we find the
following simplification of (\ref{eqf}):
 \beq\label{eqf2}z\ep\pa_{z}f + \frac G2
  \bar{f}(\ep\pa_{z})[f^2(z/2)]-(\mu-\ep)
f=0.\eeq

2. Now we assume that $f$ is polynomial of degree $n$ and a
solution of (\ref{eqf2}). We are going to show that there is a
contradiction due to the term of highest degree in the equation.
Indeed, if $f$ is a polynomial of degree $n$, then
 $(\ep\pa_{z})^k[f^2(z/2)]$ is of degree $2n-k$. But (\ref{eqf2})
 implies that
  $\bar{f}(\ep\pa_{z})[f^2(z/2)]$ is of degree $n$, hence $f$ must
  be equal to $cz^n$. This is indeed a solution of (\ref{eqf2}) if
  $n\ep+G|c|^2\ep^n(2n)!/(2^{2n+1}n!)-\mu+\ep=0$. Using
  that $\int |f|^2 e^{-|z|^2/\ep}
  =1$, we find that $|c|^2\pi\ep^{n+1}n!=1$.
  The Stirling formula provides the existence of a constant $c_0$
  such that $n\epsilon + {c_0 G}/({2\pi \epsilon}\sqrt n) \leq
  \mu$.
  For the minimizer, $\mu$ is of the same order
  as the energy, thus of order 1, so that if
   $\ep$ is too small, no $n$ can satisfy this last identity hence
  the minimizer is not a polynomial.  A
  similar argument can be used to check that, if $f$ is more
  generally a holomorphic function in ${\cal F}$,
   then it cannot have a finite
  number of zeroes. The detailed proof will be given in
  \cite{ABN}.

{\bf Approximation of a slow varying profile by the LLL} {\em The
Abrikosov solution} The Abrikosov problem \cite{A} consists in
minimizing the ratio $\left < |u|^4\right >/\left < |u|^2\right
>^2$ over periodic functions,
 where $\left < . \right
>$ denotes the average value over a cell, for functions $u$ obtained
 as limits of LLL functions. The minimum is achieved for
  $u=u_\ep(z,e^{2i\pi/3})$ where \cite{Tk}
\begin{equation}
  \label{eq.Abriko}u_\ep(z,\tau)=e^{-{|z|^{2}/2\ep}}f_\ep(z,\tau),\
  f_\ep(z,\tau)=e^{ z^{2}/2\ep}\Theta(\sqrt{\frac{\tau_{I}}{\pi
  \ep}}z, \tau)\end{equation}
  and for any complex number $\tau=\tau_R+i\tau_I$,
\beq\label{theta}
\Theta(v,\tau)=\frac{1}{i}\sum_{n=-\infty}^{+\infty}(-1)^{n}e^{i\pi\tau(n+1/2)^{2}}
e^{(2n+1)\pi iv}. \eeq The $\Theta$ function has the following
properties
\beq\label{thetaprop}\Theta(v+k+l\tau,\tau)=(-1)^{k+l}e^{-2i\pi
lv} e^{-i\pi l\tau}\Theta(v,\tau)\eeq so that $|u_\ep(z,\tau)|$ is
periodic over the lattice $\sqrt{\frac{\pi\ep}{\tau_I}}\zzz\oplus
\sqrt{\frac{\pi\ep}{\tau_I}}\zzz\tau$, and vanishes at each point
of the lattice. Without loss of generality, one can restrict
$\tau$ to vary in $|\tau|\geq 1$, $-1/2\leq \tau_R<1/2$: this is
equivalent to require that the smallest period for $\Theta$ is 1
and along the $x$ axis (see  \cite{KC}) and any lattice in the
plane, can be obtained from one of these by similarity.

For any $\tau$, $f_\ep$ given by (\ref{eq.Abriko}) is a solution
of \beq\label{eqfabri}\Pi(|f_\ep|^2e^{-|z|^2/\ep}f_\ep)=\la_\tau
f_\ep,\hbox{ with } \la_\tau=\left < |u_\ep|^2\right >
b(\tau),\eeq and \beq\label{btau}b(\tau)=\frac {\left <
|u_\ep|^4\right >}{\left < |u_\ep|^2\right
>^2}=\sum_{k,l\in \zzz} e^{-\pi |k\tau -l|^2/\tau_I}
.\eeq This expression can be obtained using arguments in \cite{T}.
The minimal value of $b(\tau)\sim 1.16$ is achieved for
$\tau=e^{2i\pi/3}$, that is for the triangular lattice \cite{K}:
in \cite{K}, it is argued that one can restrict to $\tau_R=-1/2$,
 and vary $\tau_I$ in $(1/2,\sqrt3/2)$. Accepting this restriction, they
compute the variations of $b$ which depends on a single parameter
and is indeed minimal for the triangular lattice. In \cite{ABN},
we prove that this restriction is rigorous using the description
 of these lattices by varying $\tau$ for $|\tau|=1$ and $\tau_R\in
(-1/2,0)$.

If one compares (\ref{eqfabri}) and (\ref{eqf}), one notices that
they only differ by the term $\Pi(|z|^2f)=\ep z\partial_z f+\ep
f$, which is negligible on the lattice size, but plays a role
 on the shape of the density profile.

{\em The role of the confining potential} A natural candidate to
approximate any slow varying profile $\a (z,\bar z)$ is to take
$\a(z,\bar z)u_\ep(z,\tau)$, where $u_\ep$ is the periodic
function defined in (\ref{eq.Abriko}). Of course, such a function
is not in the LLL, but  can be well approximated in the LLL by
$f^\a e^{-{|z|^{2}/2\ep}}$ where $f^\a=\Pi(\a f_\ep)$, $\Pi$ is
the projector onto the LLL (\ref{pi}) and $f_\ep$ comes from
(\ref{eq.Abriko}).
 Estimating the energy of $f^\a$ yields
 $E(f^\a)-\int  [
 |z|^2|\a|^2 \left < |u_\ep|^2\right
>+\frac {Gb(\tau)}2 |\a|^4 \left < |u_\ep|^2\right >^2 ]d^2r$
 $\sim C\ep^{1/4}.$
This computation uses calculus on $\Pi$ \cite{ABN}, and
 that
$u_\ep$ and $\a$ do not vary on the same
 scale, hence the integrals can be decoupled. The contribution of $u_\ep$
  to the energy is through the coefficient $b(\tau)$, which is minimum
 for $\tau=e^{2i\pi/3}$.

 Using pseudo differential calculus, one can show \cite{ABN},
  when $\ep$ is
small, that $f^\a$ is
 very close to $\a u_\ep$: the error is at most like
  $\ep^{1/4}$ if $\a$ is not more singular than an inverted
  parabola.
   In particular,  when $\a$ is an inverted parabola,
 this implies that in the
 Thomas Fermi region, the distribution of visible vortices is almost that
 of the triangular lattice since $\a u_\ep$ is a good
 approximation. Outside the support of the inverted parabola,
 where $f^\a$
 is very small,
  one can check that the density of distribution of zeroes of
  $f_\a$ decreases like $1/|z|$ for large $|z|$. Contrary to what
  was explained in \cite{WBP,S}, it is not a small distortion of
  the lattice which results in large changes in the density
  distribution, but a very specific and far from uniform
  distribution of the invisible
  vortices (outside the Thomas Fermi region)
  which allows to approximate an inverted parabola.

The special shape of the inverted parabola comes out if one
 wants to approximate the equation of the minimizer of the energy:
 for any $\lambda$, we can prove that
 \begin{align}\nonumber&\Pi\Bigl (
(|z|^2+G|f^\a|^2e^{-|z|^2/\ep}-\lambda)f^\a\Bigr
)+C\ep^{1/4}\\&\sim\Pi\Bigl ( (|z|^2+Gb(\tau)\left <
|u_\ep|^2\right
> |\a|^2-\lambda)\a f_\ep\Bigr )\label{simp}\end{align}
where $C$ only depends on bounds on $\a$. In other words, in the
equation for $f^\a$, one can separate in the term
$|f^\a|^2e^{-|z|^2/\ep}$
 the contributions due to the lattice and to the profile. The
right hand side of
 (\ref{simp}) is zero if $\a$ is the
inverted parabola
$$ \a(z)= \sqrt{\frac{2}{\pi
  R_0^{2}\left < |u_\ep|^2\right
>}\Bigl (1-\frac{|z|^{2}}{R_0^{2}}\Bigr)},\
  R_0=\left(\frac{2Gb(\tau)}{\pi}
\right)^{1/4}$$ and $\la=R_0^2$, so that
 $f^\a$ is almost a solution of (\ref{eqf}), up to an error in $\ep^{1/4}$.

{\em Variations of the lattice} This approach can be used to study
the variations in energy
 due to deformations of the lattice. The triangular lattice,
 corresponding to $\tau^1=e^{2i\pi/3}$, is such that the Hessian of
 $b(\tau)$ is isotropic ($\sim 0.68 Id$).
 Two
 lattices
   close to each other can be described by two close complex numbers
   $\tau^1$ and
    $\tau^2$;  the difference in
   energy between $E(f^\a(.,\tau^1))$ and $E(f^\a(.,\tau^2))$ is
   at leading order
$$\frac G4 \frac{\partial^2 b}{\partial \tau_R^2}
|\tau^1-\tau^2|^2\int |\a|^4 \left < |u_\ep|^2\right >^2d^2r\sim
\frac {0.68G}{3\pi R_0^2}|\tau^1-\tau^2|^2$$
  This computation justifies the approach
 which consists in decoupling the lattice contribution from the
 profile contribution in the energy
  \cite{S} but, given the definition of $f^\a$ using $\Pi$, it
   relies on strong deformations of the lattice for
  points far away from the Thomas Fermi region.
 For a shear deformation for which
 $u_{ij}$ are the components of the deformation tensor,
  $\tau^2-\tau^1=i\sqrt 3 u_{xy}$.
 The elastic coefficient $C_2$ is defined by the fact that the
difference in energy should be $4C_2 u_{xy}^2$. This separation of
scales allows to compute $C_2\sim 0.68 G/(4 \pi R_{0}^{2})$ (see
also \cite{S,Sinova02}) and relate it in BEC to the same one
computed for the Abrikosov solution.

{\bf Approximation by  polynomials and modes} An interesting
issue, especially for the computations of modes, is to  get
 an estimate of the degree of the polynomial which could approximate $f^\a$,
 since this function has an infinite number of zeroes. We can prove
  \cite{ABN} that
 as the degree of the polynomial gets large, the minimum of the energy for
 the problem restricted to polynomials (and the computation of modes) is a
 good approximation of the full problem. The convergence rates that we
 obtain are not satisfactory yet. We believe that a good degree should be
 $\kappa /\ep$, with $\kappa>R_0^2$ and $R_0$ is the radius
  of the inverted parabola. Given
 that the volume of the cell is $\pi \ep$, $\kappa=R_0^2$ would correspond
 to having only the visible vortices. Numerical simulations indicate that a
 sufficient number of invisible vortices is needed to recreate the inverted
 parabola profile \cite{ABD}. There are two types of invisible vortices:
 those close to the boundary of the inverted parabola which contribute to
 the bulk modes and those sufficiently far away which produce single
 particle excitations as explained in
 \cite{Macdo}. An open issue is to understand the location of these latter
 invisible vortices; some simulations suggest
  that they lie on concentric circles, but then the density of these
   circles should be very low to match our predicted global vortex density
   far away which behaves like $1/|z|$.
\begin{figure}
 \centerline{\includegraphics[width=6.5cm]{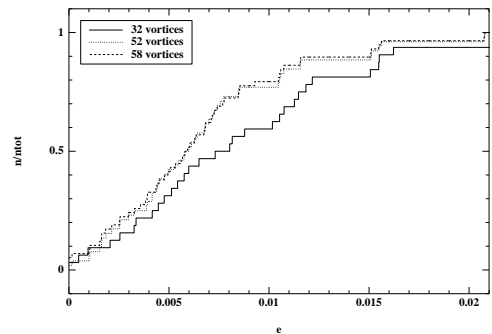}}
 \caption{Number of modes $n/n_{tot}$ having a lower energy
  than a given energy $e$ for $G=3$, $\O=0.999$}
 \label{fig}
\end{figure}
 We have performed numerical simulations with
   $\O=0.999$ and $G=3$: this fixes the number
    of visible vortices to 30, and we vary the number of
 total vortices $N$. One needs at
 least $N=52$ (that is 22 invisible vortices)
  to properly approximate the inverted parabola, the energy minimizer and the bulk
  modes.
   The distortion
 of the lattice is small at the edges but large at large
 distances. For $N$ too small, some modes do not appear
  (see Figure \ref{fig}), while
 for $N$ very large, one expects higher modes that \cite{Macdo,Macdo2}
 interpret as single particle modes.

{\bf Conclusion}
 We have shown that for the minimizer of the Gross Pitaevskii
 energy in the LLL,
the lattice of vortices is infinite, but not uniform. Any slow
varying profile can be approximated in the LLL by distorting the
lattice. This is proved using an explicit expression for the
projection onto the LLL. Our results also give an insight on the
elastic coefficient $C_2$ and the approximation of the minimizer
and modes by polynomials.

{\em Acknowledgements: } We are very indebted to Jean Dalibard and
Allan MacDonald for stimulating  discussions. Part of them took
place at the "Fondation des Treilles" in Tourtour which  hosted a
very fruitful interdisciplinary maths-physics meeting on these
topics. We thank James Anglin, Sandy Fetter and Sandro Stringari
for interesting comments.

\end{document}